\documentclass[12pt]{article}  
\usepackage{amsmath}
\usepackage{amsfonts,amssymb}

\newcommand{\cH}{{\mathcal H}}

\newcommand{\cM}{{\mathcal M}}

\newcommand{\cP}{{\mathcal P}}
\newcommand{\cE}{{\mathcal E}}

\newcommand{\cL}{{\mathcal L}}

\def\R{{\mathbb R}}
\def\N{{\mathbb N}}
\def\Z{{\mathbb Z}}
\def\C{{\mathbb C}}

\newcommand{\1}{{\bf 1}}

\newtheorem{definition}{Definition}[section]
\newtheorem{theorem}{Theorem}[section]

\newtheorem{proposition}{Proposition}[section]
\newtheorem{lemma}{Lemma}[section]

\newtheorem{remark}{Remark}[section]

\newenvironment{proof}{\bigskip\par\noindent{\it Proof:}}{$\square$\newline\vspace*{0.2cm}}

\makeatletter
    
    \@addtoreset{equation}{section}
\makeatother

\makeatother 
\begin{document}

\title{{Remarks on the Structure of Dirichlet Forms on Standard
Forms of von Neumann Algebras}}

\author{Yong Moon Park}
\date{
  { \small Department of Mathematics and Institute for Mathematical Sciences,\\
  Yonsei University, Seoul 120-749, Korea \\
 E-mail : ympark@yonsei.ac.kr }}



\maketitle
\begin{abstract}
For a von Neumann algebra $\cM$ acting on a Hilbert space $\cH$
with a cyclic and separating vector $\xi_0$, we investigate the
structure of Dirichlet forms on the natural standard form
associated with the pair $(\cM, \xi_0)$. For a general Lindblad
type generator $L$ of a conservative quantum dynamical semigroup
on $\cM$, we give sufficient conditions so that the operator $H$
induced by $L$ via the symmetric embedding of $\cM$ into $\cH$ to
be self-adjoint. It turns out that the self-adjoint operator $H$
can be written in the form of a Dirichlet operator associated to a
Dirichlet form given in \cite{Par}. In order to make the
connection possible, we also extend the range of applications of
the formula in \cite{Par}.
\end{abstract}

\section{Introduction}
The purpose of this work is to investigate the structure of
Dirichlet forms on a von Neumann algebra $\cM$ acting on  a
Hilbert space $\cH$ with a cyclic and separating vector $\xi_0$.
We are looking for a connection between Lindblad type generators
of conservative quantum dynamical semigroup(q.d.s.) on $\cM$
\cite{Lin} and Dirichlet operators associated to Dirichlet forms
introduced in \cite{Par}. In order to make the connection possible
we first extend the range of applications of the formula of
Dirichlet forms in \cite{Par}. We then consider a general Lindblad
type generator $L$ of a conservative q.d.s. on $\cM$. We give
sufficient conditions under which the operator $H$ induced by $L$
via the symmetric embedding of $\cM$ into $\cH$ is self-adjoint.
It turns out that the self-adjoint operator $H$ can be expressed
in the form of a Dirichlet operator associated to a Dirichlet form
given in \cite{Par}. In this sense, the Dirichlet forms
constructed in \cite{Par} can be considered to be natural.

The need to construct Markovian semigroups on von Neumann
algebras, which are symmetric with respect to a non-tracial state,
is clear for various applications to open systems\cite{Dav},
quantum statistical mechanics\cite{BR} and quantum
probability\cite{Acc,AFL,Part}, Although on the abstract level we
have quite well-developed theory\cite{Cip,GL1,GL2}, the progress
in concrete applications is very slow. One of the reasons is that
the general structure of Dirichlet forms for non-tracial states is
not well-understood compared to the tracial
case\cite{AG,AHO,Arv,CS}. For constructions of Dirichlet forms for
non-tracial states, we refer to
\cite{BKP1,BKP2,CFL,KP,MZ1,MZ2,QSV,Par} and the references there
in.
 In \cite{Par}, we gave a
general construction method of Dirichlet forms on standard forms
of von Neumann algebras.  The method has been used to construct
(translation invariant) symmetric Markovian semigroups for quantum
spin systems\cite{Par}, the CCR and CAR algebras with respect to
quasi-free states\cite{BKP1,BKP2} and quantum mechanical
systems\cite{BK}.

Let us describe the content of this paper briefly. Let $\cM$ be a
$\sigma$-finite von Neumann algebra acting on a Hilbert space
$\cH$ with a cyclic and separating vector $\xi_0$ for $\cM$. Let
$\Delta$ and $J$ be the modular operator and modular conjugation
respectively associated with the pair $(\cM, \xi_0)$\cite{BR}.
Denote by $\sigma_t$, $t\in \R$, the group of modular
automorphisms : $\sigma_t(A) =\Delta^{it} A\Delta^{-it}$, $A\in
\cM$. The map $j: \cM \to \cM'$ is the antilinear $*$-isomorphisms
defined by $j(A) :=JAJ$, $A\in \cM$, where $\cM'$ denotes the
commutant of $\cM$. For any $\lambda >0$, denote by
$\cM_{\lambda}$ the dense subset of $\cM$ consisting every
$\alpha_t$-analytic element of $\cM$ with a domain containing the
strip $\{ z \in \C : | Im \,z | \le \lambda \}$.

As mentioned before, we are looking for a connection between
Lindblad type generators and Dirichlet operators associated to
Dirichlet forms constructed in \cite{Par}. To make a connection
possible, we need to extend the range of applications of the
formula given in \cite{Par}. In \cite{Par}, we constructed a
Dirichlet form for any $x \in \cM_{1/4}$ and admissible function
$f$ [23, Theorem 3.1]. In this paper, we consider the function
$f_0 : \R \to \R$ given by
\begin{equation} \label{*1.1}
f_0(t) = 2(e^{2\pi t} + e^{-2\pi t} ) ^{-1}.
\end{equation}
The function $f_0$ will play a special role. We extend the
construction of Dirichlet forms to the function $f_0$ in Theorem
\ref{thm2.1}.

We next consider the generators of conservative q.d.s. on $\cM$.
The most natural generator would be the following Lindblad type
generator \cite{Lin,Part} :
\begin{equation}\label{*1.2}
L(A) = \sum_{k=1}^\infty \left\{ y_k^* y_k A -2 y_k^* A y_k + A
y_k^* y_k \right \} + i [Q,A], \quad A \in \cM,
\end{equation}
where $y_k \in \cM, \, k \in \N$, and $\sum_{k=1}^\infty y_k^*
y_k$ converges strongly. Here we have used the notation $[A,B] :=
AB-BA,\,  \forall A, B\in \cM$. However, in order to avoid the
convergence problems in the study of (\ref{*1.2})(see Remark
\ref{rem2.2} (c)), we concentrate to the case in which only finite
$y_k$'s in (\ref{*1.2}) are not zero.

For given $\{ y_1, y_2, \cdots, y_n \} \subset \cM_{1/2}$ and
$Q=Q^* \in \cM_{1/2}$, we consider the following Lindblad type
generator $L$ of a conservative q.d.s. :
\begin{eqnarray}
L&:& \cM \to \cM , \nonumber \\
L(A) &=& \sum_{k=1} ^ ny_k^*y_k A -2 y_k^* Ay_k + Ay_k^* y_k +
i[Q, A], \quad A \in \cM. \label{in1.3}
\end{eqnarray}
 Consider the
following symmetric embedding \cite{Cip}:
\begin{eqnarray*}
i_0 &:& \cM \to \cH, \\
i_0 (A) &= & \Delta^{1/4} A \xi_0 , \,\, A \in \cM,
\end{eqnarray*}
and define the operator $H$ on $\cH$ by
\begin{equation}\label{*1.3}
H\Delta^{1/4} A \xi_0 = \Delta^{1/4} L(A) \xi_0, \quad A \in \cM.
\end{equation}
If $H$ is self-adjoint, $H$ generates a symmetric Markovian
semigroup on $\cH$\cite{Cip}.

Let $L : \cM \to \cM$ be given as (\ref{in1.3}).
 Put $x_k := \sigma_{i/4}(y_k),\,\, k =1,2,\cdots, n$. Assume
that the following property holds:
\begin{equation} \label{*1.4}
\sum_{k=1} ^n x_k j(x_k) =\sum_{k=1} ^n x_k ^* j(x_k^*).
\end{equation}
Then the operator $H$ associated to $L$ by the relation
(\ref{*1.3}) is self-adjoint if and only if $Q$ is given by
\begin{equation} \label{**1.5}
Q= \sum_{k=1}^ n Q_k
\end{equation}
where
\begin{equation} \label{*1.6}
Q_k= i \int \sigma_t \left(x_k ^* \sigma_{-i/2}(x_k)
-\sigma_{i/2}(x_k^*)x_k \right) f_0 (t) \,dt,
\end{equation}
where $f_0$ is the function given in (\ref{*1.1}). Moreover, under
the condition (\ref{*1.4}), the self-adjoint operator $H$ can be
written as
$$
H = \sum_{k=1} ^n H_k ,
$$
where each $H_k, \, k=1,2,\cdots, n$, is the Dirichlet operator
associated to the Dirichlet form constructed in \cite{Par} with
$x=x_k$ and $f=f_0$. See Theorem \ref{*thm2.2} for details. Thus
conditions (\ref{*1.4}) and (\ref{**1.5}) are sufficient
conditions for $H=H^*$.

In Section 5, we give a brief discussion on necessary and
sufficient conditions for $H=H^*$ and show that, if $\xi_0$
defines a tracial state: $\langle \xi_0 , AB \xi_0 \rangle =
\langle \xi_0 , BA \xi_0 \rangle$, $ \forall A, B \in \cM$, then
the conditions (\ref{*1.4}) and (\ref{**1.5}) are also necessary
conditions for $H=H^*$. Thus we believe that the conditions
(\ref{*1.4}) and (\ref{**1.5}) are very close to necessary
conditions for $H=H^*$ for any non-tracial $\xi_0$.

We organize the paper as follows: In Section 2, we introduce
notation, definitions and necessary terminologies in the theory of
noncommutative Dirichlet forms in the sense of Cipriani\cite{Cip}.
We then list our main results(Theorem \ref{thm2.1}, Proposition
\ref{*prop2.1} and Theorem \ref{*thm2.2}). We prove Theorem
\ref{thm2.1} in Section 3, and Proposition \ref{*prop2.1} and
Theorem \ref{*thm2.2} in Section 4 respectively. In Section 5, we
give a brief discussion on the necessary and sufficient conditions
for $H=H^*$, and on the map $L$ on $\cM$ associated to a Dirichlet
operator $H$ for a general  admissible function.

\section{ Notation, Definitions and Main Results}

In this section, we first  first introduce necessary terminologies
in the theory of  Dirichlet forms and Markovian semigroups on
standard form of von Neumann algebras\cite{Cip} and then list our
main  results.

Let $\cM$ be a $\sigma$-finite von Neumann algebra
 acting on a complex Hilbert space $\cH$ with an inner product $\langle \cdot , \cdot \rangle$.
 Let $\xi_0 \in \cH$ be a cyclic and separating vector
 for $\cM$. We use $\Delta$ and  $J$ to denote respectively, the modular operator and
 the modular conjugation associated with the pair $(\cM,\, \xi_0 )$\cite{BR}. The associated
 modular automorphism group is denoted by $\sigma_t$: $\sigma_t (A) = \Delta^{it} A \Delta^{-it},
 \forall A\in \mathcal{M},\,t\in\mathbb{R}$. The map
  $j:\mathcal{M}\rightarrow\mathcal{M}'$ is the antilinear
 $*$-isomorphism defined by $j(A)=JAJ,\,A\in\cM$.

 The natural positive cone $\mathcal{P}$ associated with the pair
  $(\mathcal{M}$, $\xi_0 )$ is the closure of the set
 $$
 \{ Aj(A)\xi_0 : A\in \mathcal{M}\}.
 $$
By a general result, the closed convex cone $\cP$ can be obtained
by the closure of the set
 $$
 \{\Delta^{1/4} AA^* \xi_0 : A\in \mathcal{M}\}
 $$
and this cone $\cP$ is
  self-dual  in the sense  that
$$\Big\{\xi \in \cH : \langle\xi, \eta\rangle \ge 0,\,\,\forall \eta \in \cP\Big\}=\cP.$$
For the details we refer \cite{Ara} and Section 2.5 of \cite{BR}.

   The form
$(\mathcal{M},\mathcal{H},\mathcal{P},J)$ is the standard form
associated with the pair $(\mathcal{M},\xi_0 )$.  We shall use the
fact that $\cH$ is  the complexification
 of the real  subspace
 $ \cH^J =\Big\{\xi\in\cH : \langle\xi, \eta\rangle\in\mathbb{R},\,\,\forall\eta\in\cP\Big\},$
 whose elements are called {\it $J$-real}: $\cH =\cH^J \bigoplus i \cH^J .$
  The cone $\cP$ gives rise to a structure of ordered Hilbert space on $\cH^J$
 (denoted by $\leq$) and to an anti-unitary involution $J$ on $\cH$,
 which preserves $\cP$ and $\cH^J$: $J(\xi +i\eta) = \xi -i\eta,\,
 \forall \xi, \eta \in \cH^J$.  Also note that any $J$-real element $\xi \in \cH^J$ can be decomposed uniquely
 as a difference  of two mutually orthogonal, positive elements,
 called the positive and  negative part of $\xi$, respectively : $\xi =\xi_+ - \xi_-$, $\xi_+,\,\xi_- \in \cP$
 and $\langle\xi_+, \xi_- \rangle =0$. The order interval $\{ \eta \in \cH : 0 \le \eta \le \xi_0 \}$ will
 be denoted by $[0, \xi_0 ]$.  This is a closed convex  subset of $\cH$, and we shall denote the nearest
 point projection onto $[0,\xi_0]$ by $\eta \mapsto \eta_{I}$.

A bounded operator $A$ on $\cH$ is called {\it $J$-real} if
$AJ=JA$ and {\it positive preserving} if $A\cP \subset \cP$. The
semigroup $\{T_t\}_{t\ge 0}$ is said to be {\it $J$-real} if $T_t$
is $J$-real for any $t\ge0$ and it is called {\it positive
preserving} if $T_t$ is  positive preserving for any $t\ge 0$.
 A bounded operator $A:\cH \to \cH$ is called {\it sub-Markovian}
(with respect to $\xi_0$) if $0\le \xi\le \xi_0$ implies $0\le
A\xi \le \xi_0$.  $A$
 is called {\it Markovian} if it is sub-Markovian and also
 $A\xi_0=\xi_0$. A semigroup $\{T_t\}_{t\ge 0}$ is said to be {\it
 sub-Markovian} (with respect to $\xi_0$) if $T_t$ is sub-Markovian for
 every $t\ge 0$. The semigroup $\{ T_t\}_{t\ge 0}$ is called {\it
 Markovian} if $T_t$ is Markovian for every $t\ge 0$.

Next, we consider a sesquilinear form on some linear manifold of
$\cH$ : $\cE(\cdot,\cdot): D(\cE)\times D(\cE) \to \C$. We also
consider the associated quadratic form: $\cE[\cdot]: D(\cE)\to
\C$,  $\cE[\xi]:= \cE(\xi,\xi)$.  A real valued quadratic form
$\cE[\cdot]$ is said to be {\it semi-bounded} if $\inf \{ \cE[\xi]
: \xi \in D(\cE), \,\, ||\xi|| =1\} =-b > -\infty$. A quadratic
form $(\cE, D(\cE))$ is said to be {\it $J$-real} if $J D(\cE)
\subset D(\cE)$ and $\cE[J\xi] = \overline{\cE[\xi]}$ for any $\xi
\in D(\cE)$. For a given semi-bounded quadratic form $\cE$, one
considers the inner product given by
$\langle\xi,\eta\rangle_\lambda:= \cE(\xi,\eta) + \lambda
\langle\xi,\eta\rangle$, for $\lambda>b$. The form $\cE$ is {\it
closed} if $D(\cE)$ is a Hilbert space for some of the above
norms. The form $\cE$ is called {\it closable} if it admits a
closed extension.

Associated to a semi-bounded closed form $\cE$, there are a
self-adjoint operator $(H, D(H))$ and a strongly continuous,
symmetric semigroup $\{T_t\}_{t\ge0}$. Each of the above objects
determines uniquely the others according to well known relations
(see \cite{RS} and Section 3.1 of \cite{BR}).

A $J$-real, real-valued, densely defined quadratic form $(\cE,
D(\cE))$ is called {\it Markovian} with respect to $\xi_0\in \cP$
if
\begin{equation*}
\eta \in D(\cE)^J \text{  implies  } \eta_I \in D(\cE) \text{ and
} \cE[\eta_I] \le \cE[\eta],
\end{equation*}
where $D(\cE) ^J := D(\cE) \cap \cH^J$. A closed Markovian form is
called a {\it Dirichlet form}.

Next, we collect main results of \cite{Cip}. Let $(\cE, D(\cE))$
be a $J$-real, real valued, densely defined closed form. Assume
that the following properties hold:
\begin{eqnarray}\label{2.1}
             &(a)& \xi_0 \in D(\cE), \\
             &(b)&  \cE(\xi,\xi)\ge0 \,\,\text{for} \,\,\xi\in D(\cE), \nonumber \\
             &(c)& \xi\in D(\cE)^J \,\text{ implies } \, \xi_{\pm} \in D(\cE) \,\,\text{
             and}\,\,
 \cE(\xi_+ , \xi_{-} )\leq 0.\nonumber
    \end{eqnarray}
    Then $\cE$ is a Dirichlet form if and only if
    $\cE(\xi,\xi_0)\ge0$ for all $\xi\in D(\cE)\cap\cP$.
 The above result follows from Proposition 4.5 (b) and Proposition
4.10 (ii) of \cite{Cip}.

The following is Theorem 4.11 of \cite{Cip} :
 Let $\{T_t \}_{t\ge 0}$ be a $J$-real, strongly continuous, symmetric semigroup on $\cH$
 and let $(\cE, D(\cE))$ be the  associated densely defined $J$-real, real
 valued quadratic form. Then the followings are equivalent.
 \begin{eqnarray} \label{2.2}
             &(a)& \{ T_t \}_{t\ge 0} \,\, \text{ is
             sub-Markovian}.\\ \nonumber
             &(b)& (\cE, D(\cE))\,\,\text{ is a Dirichlet form .}
    \end{eqnarray}
We refer the reader to \cite{Cip} for the details.

Next, we give an extended version of the general construction
method developed in \cite{Par}. For any $\lambda >0$, denote by
$I_\lambda$ the closed strip given by
\begin{equation} \label{2.3}
I_\lambda = \{ z : z \in \C , | Im\, z | \le \lambda \}.
\end{equation}
Let us introduce the notion of admissible functions \cite{Par}.
\begin{definition} \label{defn2.1}
An analytic function $f: D \to \mathbb{C}$ on a domain $D$
containing the strip $I_{1/4}$ is said to be admissible if the
following properties hold:
\begin{enumerate}
 \item[(a)] $f(t) \ge 0 \quad \text{for}\quad  \forall t\in
 \mathbb{R},$
\item[(b)] $f(t+i/4) + f(t-i/4) \ge 0 \quad \text{for}\quad
\forall t\in \mathbb{R},\nonumber$
 \item[(c)] there exist $M>0$ and $p>1$ such that the bound
    $$ |f(t+is)| \le M(1+|t|)^{-p} $$
   holds uniformly in $s\in [-1/4,1/4].$
 \end{enumerate}
\end{definition}
We remark that there exist a non-trivial admissible function[23,
Lemma 3.1].

Next, we consider the function $f_0 : \R \to \R$ given by
\begin{equation}\label{2.4}
f_0 (t) = 2 ( e^{2\pi t } + e^{-2 \pi t } ) ^{-1} .
\end{equation}
The function $f_0$ will play important roles in the sequels. Using
the residue integration method, it is easy to check that
\begin{equation} \label{2.5}
2 \int ( e^{2 \pi t} + e^{-2\pi t} )^{-1} e^{ikt} \, dt = (e^{k/4}
+ e^{-k/4} )^{-1}.
\end{equation}
See also the expression in P. 94 of \cite{BR}. One can see that
$f_0$ has an analytic extension, denoted by $f_0$ again, to the
interior of $I_{1/4}$. On the boundary of $ I_{1/4}$, it defines a
distribution, and satisfies the equality
$$
f(t+ i/4) + f(t-i/4) = \delta(t)
$$
in the sense of distribution. Thus even if $f_0$ is not an
admissible function, it is almost admissible.

For any $\lambda >0$, denote by $\cM_\lambda$ the dense subset of
$\cM$ consisting of every $\sigma_t$-analytic element with a
domain containing $I_\lambda$. By Proposition 2.5.21 of \cite{BR},
any $A \in \cM_\lambda$ is strongly analytic. In the following,
the inner product $\langle \cdot, \cdot \rangle$ on $\cH$ is
conjugate linear in the first and linear in the second variable.
For  given $x \in \cM_{1/4}$ and an admissible function $f$ or
else $f=f_0$,  define a sesquilinear form  $\cE
:\cH\times\cH\longrightarrow\mathbb{C}$ by
\begin{eqnarray}\label{2.6}
     &&\cE(\eta,\xi)\\
     &&=\int \Big\langle\big(\sigma_{t-i/4}(x)-j(\sigma_{t-i/4}(x^* ))\big)\eta,
          \big(\sigma_{t-i/4}(x)-j(\sigma_{t-i/4}(x^* ))\big)\xi\Big\rangle f(t)dt\nonumber\\
     &&\quad +\int\Big\langle\big(\sigma_{t-i/4}(x^* )-j(\sigma_{t-i/4}(x))\big)\eta,
              \big(\sigma_{t-i/4}(x^*
              )-j(\sigma_{t-i/4}(x))\big)\xi\Big\rangle f(t)dt\nonumber
\end{eqnarray}
The form is positive and bounded. The self-adjoint operator $H$
associated to the form is given by
\begin{eqnarray}\label{2.7}
H &=& \int \left( \sigma_{t+i/4} (x^*) -j(\sigma_{t+i/4}(x))
\right) \left( \sigma_{t-i/4} (x) -j(\sigma_{t-i/4}(x^*))
\right)f(t) \,
dt \\
&&\,\,\,\, \int \left( \sigma_{t+i/4} (x) -j(\sigma_{t+i/4}(x^*))
\right) \left( \sigma_{t-i/4} (x^*) -j(\sigma_{t-i/4}(x))
\right)f(t) \, dt \nonumber
\end{eqnarray}
The following result is an extended version of Theorem 3.1 of
\cite{Par}.
\begin{theorem}\label{thm2.1}
Let $f$ be either an admissible function or else $f=f_0$ and
$x\in\cM_{1/4}$. Let $(\cE,\cH)$ be the quadratic form associated
to the sesquilinear form  defined as in (\ref{2.6}) : $\cE [\xi] =
\cE(\xi, \xi)$. Let $H$ be the self-adjoint operator associated
with $(\cE,\cH)$.
 Then the following properties
 hold:

(a) $H\xi_0 =0$,

(b) $\cE$ is $J$-real

(c) $\cE(\xi_+ ,\xi_- ) \le 0$ \,  \,$\forall \xi\in\cH^J .$\\
Furthermore the form $(\cE,\cH)$ is a Dirichlet form.
\end{theorem}

The proof of the theorem will be given in the next section. It may
be worth to compare Theorem 3.1 of \cite{Par} and Theorem 2.1 in
the above, and give a comment on possible extensions of Theorem
\ref{thm2.1}.
\begin{remark} \label{rem2.1} (a) In \cite{Par}, the properties (a),(b) and (c) in
Theorem \ref{thm2.1} were proved under assumptions that $x=x^* \in
\cM$, $f$ is admissible and that there exist a constant $M >0$
such that  the bound
$$
\sup_{s \in [-1/4, 1/4] } \| \sigma_{t+is}(x) \| \le M
$$
holds uniformly in $t \in \R$. Notice that if one writes
\begin{equation}\label{2.8}
x_1 := \frac 1 {\sqrt{2}} (x+ x^*), \quad x_2 := \frac i
{\sqrt{2}} (x-x^*),
\end{equation}
then $\cE(\eta, \xi)$ can be written as
\begin{equation} \label{2.9}
\cE(\eta , \xi ) = \frac12 \{\cE_1 (\eta, \xi) + \cE_2( \eta,
\xi)\} ,
\end{equation}
where $\cE_j( \eta, \xi)$, $j=1,2$ is the form corresponding to
$x_j$, $i=1,2$, respectively. Thus one may assume that $x$ is
self-adjoint. On the other hand, the above bound  need to prove
the property (c) by Cauchy's integral theorem. We will give
another proof of the property (c) which do not use the above
bound.

(b) In applications, one may choose $\{x_k\}_{k=1}^\infty \in
\cM_{1/4}$ which generates $\cM$, and $f$, where $f$ is an
admissible function or else $f=f_0$. For each $k \in \N$, let
$(\cE_k, \cH)$ be the Dirichlet form obtained from (\ref{2.6})
with $x=x_k$. Let $(\cE, D(\cE))$ be the sesquilinear form defined
by
\begin{eqnarray*}
D(\cE) & = & \{ \xi \in \cH : \sum_{k=1} ^\infty  \cE_k (\xi, \xi
) < \infty \}, \\
\cE(\eta, \xi )& =& \sum_{k=1}^\infty  \cE_k (\eta, \xi), \quad
\eta, \xi \in D(\cE).
\end{eqnarray*}
If $D(\cE)$ is dense in $\cH$, then $(\cE, D(\cE))$ is a Dirichlet
form [13, Theorem 5.2]. The above method has been used in
\cite{BKP1,BKP2}. Also it may be possible to extend Theorem
\ref{thm2.1} to the case in which $x$ is an unbounded operator
affiliated with $\cM$ \cite{BKP1,BK}.
\end{remark}

As discussed in Introduction, we will consider Lindblad type
generators of conservative q.d.s. on $\cM$ and their symmetric
embeddings. Recall that $\cM_{1/2}$ denotes the dense subset of
$\cM$ consisting of every $\sigma_t$-analytic element with a
domain containing the strip $I_{1/2}$. For given $y \in \cM_{1/2}$
and $Q= Q^* \in \cM_{1/2}$, we first consider the following
Lindblad type generator of a q.d.s. on $\cM$ :
\begin{eqnarray}
\nonumber L &:& \cM \to \cM \\
L(A) &=& y^* y A - 2 y^* A y + Ay^* y+ i[Q,A], \quad A \in \cM,
\label{2.10}
\end{eqnarray}
where $[A,B]:= AB-BA$, $ A, B \in \cM$. Consider the following
symmetric embedding \cite{Cip} :
\begin{eqnarray}
\nonumber i_0 &:& \cM \to \cH\\
i_0(A) &=& \Delta^{1/4} A \xi_0, \quad A \in \cM, \label{2.11}
\end{eqnarray}
and define the operator $H$ on $\cH$ by
\begin{equation} \label{2.12}
H \Delta^{1/4} A\xi_0 = \Delta^{1/4} L(A) \xi_0, \quad A \in \cM.
\end{equation}
It is easy to see that $H$ is self-adjoint if and only if $L$
satisfied the following {\it KMS symmetry} \cite{Cip,GL1} : For
any $A, B \in \cM_{1/4}$
$$
\langle \sigma_{-i/4} (L(A)) \xi_0, \sigma_{-i/4}(B) \xi_0 \rangle
=\langle \sigma_{-i/4} (A) \xi_0, \sigma_{-i/4}(L(B)) \xi_0
\rangle.
$$
According to [13, Proposition 2.4 and Theorem 2.12], the map $L$
generates a (weak* continuous) KMS symmetric, conservative q.d.s.
on $\cM$ if and only if $H$ generates a (strongly continuous)
symmetric Markovian semigroup on $\cH$.  The following result can
be considered as a structure theorem for Dirichlet forms on the
standard form $(\cM, \cH, \cP, J)$ associated to the pair $(\cM ,
\xi_0)$.

\begin{proposition} \label{*prop2.1}
For given $y \in \cM_{1/2}$ and $Q=Q^* \in \cM_{1/2}$, let $L :\cM
\to \cM$ be given as (\ref{2.10}). Put $x := \sigma_{i/4}(y)$.
Assume that the relation
\begin{equation} \label{2.13}
xj(x) = x^* j(x^*)
\end{equation}
holds. Let $H$ be the operator on $\cH$ defined as (\ref{2.12}).
Then $H$ is self-adjoint if and only if $Q$ is given by
\begin{equation} \label{2.14}
Q= i \int \left( \sigma_t(x^*)\sigma_{t-i/2}(x) - \sigma_{t+i/2}
(x^*) \sigma_t (x) \right ) f_0 (t) \,dt,
\end{equation}
where $f_0$ is the function given in (\ref{2.4}). Moreover the
self-adjoint operator $H$ can be expresses in the form of the
Dirichlet operator given as (\ref{2.7}) with $f=f_0$.
\end{proposition}

The proof of the above Proposition will be produced in Section 4.
The following result is a generalization of Proposition
\ref{*prop2.1}.

\begin{theorem}\label{*thm2.2}
Let $y_k$, $k=1,2,\cdots, n$, be elements of $\cM_{1/2}$, and
$Q=Q^* \in \cM_{1/2}$. Let $L$ be the map of $\cM$ into itself
given by
\begin{equation}\label{**2.15}
L(A) = \sum_{k=1}^n \left( y_k^* A y_k -2 y_l^* Ay_k + Ay_k^* y_k
\right) + i [Q,A], \quad A\in \cM.
\end{equation}
Put $x_k := \sigma_{i/4}(y) $, $ k=1,2,\cdots,n$. Assume that the
relation
\begin{equation} \label{**2.16}
\sum_{k=1}^n x_k j(x_k) = \sum_{k=1}^n x_k^* j(x_k^*)
\end{equation}
holds. Let $H$ be the operator on $\cH$ defined as (\ref{2.12}).
Then $H$ is self-adjoint if and only if $Q$ is given by
\begin{equation}\label{**2.17}
Q= \sum_{k=1} ^n Q_k
\end{equation}
where each $Q_k$, $k=1,2,\cdots, n$, is given as  in (\ref{2.14})
with $x$ replaced by $x_k$. Moreover the self-adjoint operator can
be written as
$$
H= \sum_{k=1}^n H_k,
$$
where each $H_k$, $k=1,2,\cdots ,n$, is given as (\ref{2.7}) with
$x=x_k$ and $f=f_0$.
\end{theorem}

We give  comments on the conditions (\ref{**2.16}) and its
consequences and possible extension of Theorem \ref{*thm2.2}  to
the general Lindblad type generator given in (\ref{*1.2}):
\begin{remark} \label{rem2.2} (a) The conditions (\ref{**2.16})
and (\ref{**2.17}) are sufficient conditions for the map $L$ given
in (\ref{**2.15}) to be KMS symmetric, or equivalently the
operator $H$ induced by $L$ to be self-adjoint. If $\xi_0$ is
tracial : $\langle \xi_0, AB \xi_0 \rangle = \langle \xi_0, BA
\xi_0 \rangle$, $\forall A, B \in \cM$, then the conditions
(\ref{**2.16}) and (\ref{**2.17}) are also necessary conditions
for $L$ to be (KMS) symmetric. See Section 5.

(b) This condition (\ref{**2.16}) is equivalent to the following
condition:
\begin{equation} \label{**2.18}
\sum_{k=1}^n \sigma_{i/4} (x_k^* )A \sigma_{-i/4} (x_k )=
\sum_{k=1}^n \sigma_{i/4} (x_k )A \sigma_{-i/4} (x_k^* ), \quad
\forall A \in \cM.
\end{equation}
See Lemma \ref{lemm4.1} (b). In terms of $x_k$'s, $L(A)$ in
(\ref{**2.15}) is given by
\begin{eqnarray*}
L(A) &=& \sum_{k=1} ^n \{ \sigma_{i/4} (x_k^* )\sigma_{-i/4} (x_k
)A -2 \sigma_{i/4} (x_k^* )A \sigma_{-i/4} (x_k )+ A \sigma_{i/4}
(x_k^* ) \sigma_{-i/4} (x_k )\} \\
&& \,\,\, + i [Q, A].
\end{eqnarray*}
For given $\{ x_1, x_2, \cdots, x_n \} \subset \cM_{1/4}$, let $\{
\widetilde{x}_1,\widetilde{x}_2, \cdots, \widetilde{x}_{2n} \}
\subset \cM_{1/4}$ be the family of self-adjoint elements defined
by
$$
\widetilde{x}_{2k} = \frac1{\sqrt{2}} (x_k + x_k^* ), \quad
\widetilde{x}_{2k-1} = \frac i{\sqrt{2}} (x_k - x_k^* ), \,\,\,
k=1,2,\cdots, n.
$$
Then, under the condition (\ref{**2.18}), the KMS symmetric map
$L$ can be written as
\begin{equation} \label{**2.19}
L(A) = \frac12 \sum_{k=1} ^{2n} L_k (A) ,
\end{equation}
where for $k=1,2, \cdots, 2n $,
\begin{eqnarray}\nonumber
L_k(A) &=& \sigma_{i/4} (\widetilde{x}_k)\sigma_{-i/4}
(\widetilde{x}_k)A -2 \sigma_{i/4} (\widetilde{x}_k)A
\sigma_{-i/4} (\widetilde{x}_k)+A \sigma_{i/4}
(\widetilde{x}_k)\sigma_{-i/4} (\widetilde{x}_k) \\
 && \,\, + i [Q_k, A], \label{**2.20}
\end{eqnarray}
where $Q_k$ is given by (\ref{2.14}) with $x=\widetilde{x}_k$.

(c) For any family $\{ x_k \}_{k=1}^\infty \subset \cM_{1/4}$ of
self-adjoint elements, consider the following Lindblad type
generator
\begin{equation} \label{2.21}
L(A) = \sum_{k=1}^\infty L_k (A), \quad A \in \cM ,
\end{equation}
where each $L_k (A), \, k \in \N$, is given by (\ref{**2.20}) with
$x_k$ replacing $\widetilde{x}_k$. Since each $L_k$, $k\in \N$, is
KMS symmetric, the map $L$ given above is formally KMS symmetric,
and the operator $H$ induced by $L$ is given by
\begin{equation} \label{2.22}
H = \sum_{k=1}^\infty H_k
\end{equation}
where each $H_k$ is the Dirichlet operator given by (\ref{2.7})
with $f=f_0$ and  $x_k$ replacing $x$. The expressiions in
(\ref{2.21}) and (\ref{2.22}) are still formal. In order to give
rigorous meanings to the expressions, one has to give dense
domains $D(L)$ and $D(H)$ such that the right hand sides of
(\ref{2.21}) and (\ref{2.22}) are well-defined. Since $Q_k$ and
$H_k$ , $k \in \N$, are given by integral forms as in (\ref{2.14})
and (\ref{2.7}) respectively, the task would not so simple. It
would be very nice if one can give a sufficient condition on $\{
x_k \}_{k=1}^\infty$, which is easy to verify for concrete models,
such that the right hand sides of (\ref{2.21}) and (\ref{2.22})
converge in a appropriate sense. See Remark \ref{rem2.1} (b).
\end{remark}

\section{Proof of Theorem \ref{thm2.1}}

Before proving Theorem \ref{thm2.1}, let us introduce linear maps
on $\cL(\cH)$ which will be used frequently in the sequels. For
any $\lambda >0$, denote by $\cL_{\lambda}(\cH)$ the dense subset
of $\cL(\cH)$ consisting of every $\sigma_t$-analytic element of
$\cL(\cH)$ with a domain containing the strip $I_{\lambda}$. Let
$D_{1/4}$ and $D_{-1/4}$ be the linear maps on $\cL(\cH)$ defined
by
\begin{eqnarray}\nonumber
D(D_{1/4}) &=& \cL _{1/4} (\cH), \\
\label{3.1} D_{1/4} (A) &=& \sigma_{-i/4}(A) , \quad A \in
\cL_{1/4} (\cH),
\end{eqnarray}
and
\begin{eqnarray}\nonumber
D(D_{-1/4}) &=& \cL _{1/4} (\cH), \\
\label{3.2} D_{-1/4} (A) &=& \sigma_{i/4}(A) , \quad A \in
\cL_{1/4} (\cH).
\end{eqnarray}
Put
\begin{eqnarray}
\label{3.3} T &:= & D_{1/4} + D_{-1/4}, \\
S&:=& D_{1/4} - D_{-1/4}. \nonumber
\end{eqnarray}
Let $I_0$ be the linear map defined by
\begin{eqnarray}
\nonumber D(I_0) &= &\cL (\cH), \\
I_0 (A) &=& \int \sigma_t (A) f_0 (t) dt, \quad A \in \cL(\cH),
\label{3.4}
\end{eqnarray}
where $f_0$ is the function given in (\ref{2.4}).

We have the following result which will be used in the proofs of
the results in Section 2.
\begin{lemma} \label{lemm3.1}
The relations
$$
TI_0 (A) = I_0 T(A) =A
$$
hold for any $A\in \cL_{1/4}(\cH)$. That is, $T$ is  invertible
and $T^{-1} = I_0 $.
\end{lemma}
\begin{proof}
The proof of the above lemma is essentially contained in the proof
of Theorem 2.5.14(Tomita-Takesaki theorem) of \cite{BR}. Since the
method of the proof will be used in the proof of Theorem
\ref{thm2.1}, we produce the proof.

As a relation between bilinear forms on $D(\Delta^{1/4} ) \cap
D(\Delta^{-1/4})$, one has
$$
T(A) = \Delta^{1/4} A \Delta^{-1/4} + \Delta^{-1/4} A
\Delta^{1/4}, \quad A \in \cL (\cH).
$$
Now take $\eta, \xi \in D(\Delta^{1/4} ) \cap D(\Delta^{-1/4})$.
Then it follows that for any $A \in \cL_{1/4}(\cH)$
\begin{eqnarray}\nonumber
&&\langle \eta, T(I_0 (A)) \xi \rangle = \langle \Delta^{1/4}
\eta, I_0 (A) \Delta^{-1/4}\xi \rangle + \langle \Delta^{-1/4}
\eta, I_0 (A) \Delta^{1/4}\xi \rangle \\
\label{3.5} && \quad =\int \left( \langle \Delta^{-it + 1/4} \eta,
 A\Delta^{-it-1/4}\xi \rangle + \langle \Delta^{-it -1/4} \eta, A \Delta^{-it+1/4}\xi
 \rangle\right ) f_0 (t)\,dt.
\end{eqnarray}
Denote by $h$ the generator of $\Delta^{it}$ :
$ h :=
\log(\Delta).
$
Using the spectral decomposition of $h$:
$$ h= \int \mu \, dE(\mu),
$$
one obtains that
\begin{eqnarray*}
&&\langle \eta, T(I_0 (A)) \xi \rangle \\
&& \quad = \int f_0 (t) \left\{ \int d^2 \langle E(\mu) \eta, A
E(\rho) \xi \rangle  \left( e^{(\mu-\rho )/4} + e^{-(\mu-\rho) /4}
\right) e^{i(\mu-\rho)t} \right\}\, dt.
\end{eqnarray*}
The domain restrictions on $\eta$ and $\xi$ allow interchange of
the order of integrations and one has
\begin{eqnarray}
\label{3.6} && \langle \eta, T(I_0 (A)) \xi \rangle \\
&& \quad =\int d^2 \langle E(\mu) \eta, A E(\rho) \xi \rangle
\left( e^{(\mu-\rho)/4} + e^{-(\mu-\rho)/4} \right) \int f_0(t)
e^{i(\mu-\rho)t} \,dt \nonumber\\
&& \quad = \int d^2 \langle E(\mu)\eta, A E(\rho) \xi \rangle
\nonumber \\
&&\quad = \langle \eta, A \xi \rangle, \nonumber
\end{eqnarray}
where the first step uses the Fourier relation in (\ref{2.5}).
Thus as a relation between bilinear forms in $D(\Delta^{1/4} )
\cap D(\Delta^{-1/4})$, we have
$$
T(I_0 (A)) =A.
$$
For any $A \in \cL_{1/4} (\cH)$, $ I_0 (A) \in \cL_{1/4}(\cH)$ and
$T(I_0 (A)) \in \cL(\cH)$. Since $D(\Delta^{1/4} ) \cap
D(\Delta^{-1/4})$ is dense in $\cH$. the above equality holds as a
relation between bounded operators. It follows from the
definitions of $T$ and $I_0$ in (\ref{3.3}) and (\ref{3.4})
respectively  that $T$ and $I_0$ commute on $\cL_{1/4} (\cH)$.
This completes the proof of the lemma.
\end{proof}

We now turn to the proof of Theorem \ref{thm2.1}.

\vspace*{0.2cm} \noindent {\it Proof of Theorem \ref{thm2.1}.} The
proof of the properties (a) and (b) is same as that in the proof
of Theorem 2.1 of \cite{Par}.  See also the proof of Theorem 2.1
of \cite{BKP1}.

We prove that property (c). By the expression of $\cE(\eta, \xi)$
in (\ref{2.6}), $\cE(\xi_+, \xi_ -)$ can be written as
\begin{eqnarray}\nonumber
\cE(\xi_+, \xi_-) &=& \cE^{(1)} (\xi_+, \xi_-)+ \cE^{(2)} (\xi_+,
\xi_-) \\
&=& (\textmd{I}^{(1)} + \textmd{II}^{(1)} ) + (\textmd{I}^{(2)} +
\textmd{II} ^{(2)} ), \label{3.7}
\end{eqnarray}
where
\begin{eqnarray}
\label{3.8} \textmd{I}^{(1)} &=& \int \left( \langle
\sigma_{t-i/4}(x) \xi_+, \sigma_{t-i/4}(x) \xi_- \rangle +\langle
\sigma_{t-i/4}(x^*) \xi_-, \sigma_{t-i/4}(x^*) \xi_+ \rangle
\right) f(t)\,dt \\
\nonumber \textmd{II}^{(1)} &=& -\int \left( \langle
\sigma_{t-i/4}(x) \xi_+, j(\sigma_{t-i/4}(x^*) )\xi_- \rangle
+\langle j(\sigma_{t-i/4}(x^*)) \xi_+, \sigma_{t-i/4}(x) \xi_-
\rangle \right) f(t)\,dt,
\end{eqnarray}
and $\textmd{I}^{(2)}$ and $\textmd{II}^{(2)}$ are obtained from
$\textmd{I}^{(1)}$ and $\textmd{II}^{(1)}$, respectively,
replacing $x$ by $x^*$ in the above. Here we have used the fact
that $\langle J\eta, J \xi\rangle= \langle \xi , \eta \rangle $ in
$\textmd{I}^{(1)}$.

As a consequence of Theorem 4(7) of \cite{Ara}, $\cM\xi_+ \,\bot\,
\cM\xi_-$, which implies $\textmd{I}^{(1)} = 0$ and
$\textmd{I}^{(2)} =0$. Next, we consider $\textmd{II}^{(1)}$. It
can be checked that $\sigma_{t-is}(x)^* = \sigma_{t+is}(x^*)$ and
$j(\sigma_{t+is} (x)) = \sigma_{t-is} (j(x))$ for any $x \in
\cM_{1/2} $ and $s \in [-1/4 , 1/4]$, and so
\begin{eqnarray*}
\langle \sigma_{t-i/4} (x) \xi_+ , j(\sigma_{t-i/4}(x^*)) \xi _-
\rangle &=& \langle \xi _+, \sigma_{t+i/4}(x^*j(x^*)) \xi_-
\rangle
, \\
\langle j(\sigma_{t-i/4} (x^*)) \xi_+ , \sigma_{t-i/4}(x) \xi _-
\rangle &=& \langle \xi _+, \sigma_{t-i/4}(x j(x) )\xi_- \rangle .
\end{eqnarray*}
It follows from the definition of $\textmd{II}^{(1)}$ in
(\ref{3.8}) that
\begin{eqnarray*}
\textmd{II}^{(1)} &=& - \int \langle \xi _+, \sigma_{t+i/4}(x^*
j(x^*)
)\xi_- \rangle f(t) \,dt \\
&&  - \int \langle \xi _+, \sigma_{t-i/4}(x j(x) )\xi_- \rangle
f(t) \,dt
\end{eqnarray*}
Replacing $x$ by $x^*$ in the above, we obtain the expression of
$\textmd{II}^{(2)}$. Thus we get
\begin{eqnarray}
\nonumber \textmd{II} &= & \textmd{II}^{(1)} + \textmd{II} ^{(2)} \\
&=& - \int \langle \xi _+, T\left(\sigma_{t}(x j(x)
+x^*j(x^*))\right)\xi_- \rangle f(t) \,dt. \label{3.9}
\end{eqnarray}

We first consider the case for $f=f_0$. From the definition of
$I_0$ in (\ref{3.4}) and Lemma \ref{lemm3.1}, we have
\begin{eqnarray*}
\textmd{II} &=& - \langle \xi_+, T (I_0 (x j(x) +x^* j(x^*)))
\xi_- \rangle
\\
&=& -\langle \xi_+,  (x j(x) +x^* j(x^*)) \xi_- \rangle \\
&\le & 0.
\end{eqnarray*}
Here we have used the fact that $Aj(A) \xi_- \in \cP$ for any
$A\in \cM$.

Next we consider any admissible function $f$. For any $\eta, \zeta
\in D(\Delta^{1/4}) \cap D(\Delta^{-1/4})$, consider the following
expression:
\begin{equation} \label{3.10}
B(\eta, \zeta) := -\int \langle \eta, T(\sigma_t (xj(x) + x^*
j(x^*))) \zeta\rangle f(t) \,dt .
\end{equation}
Employing the method similar to that used to obtain the first
relation of (\ref{3.6}) from (\ref{3.5}), we have
\begin{eqnarray*}
&& B(\eta, \zeta) \\
&& \,\,\, = - \int d^2 \langle E(\mu) \eta, (x j(x) + x^* j(x^*))
E(\rho)\zeta \rangle (e^{(\mu-\rho) /4} + e^{-(\mu-\rho)/4} )
\\
&& \quad\quad\quad \cdot \int f(t) e^{i(\mu-\rho)} \,dt.
\end{eqnarray*}
We now use the bound (c) in  Definition \ref{defn2.1} and Cauchy's
integral theorem to conclude that
\begin{eqnarray*}
&&(e^{(\mu-\rho) /4} + e^{-(\mu-\rho)/4} )  \int f(t)
e^{i(\mu-\rho)} \,dt \\
&& \quad\quad \quad = \int \left( f(t-i/4) + f(t+i/4)\right)
e^{i(\mu-\rho)t} \,dt,
\end{eqnarray*}
and so
\begin{eqnarray*}
&& B(\eta, \zeta) \\
&& \,\,\, = - \int d^2 \langle E(\mu) \eta, (x j(x) + x^* j(x^*))
E(\rho) \zeta \rangle \left( f(t-i/4) +f(t+i/4) \right)
e^{i(\mu-\rho)t} \,dt \\
&& \,\,\, = - \int  \langle \eta, \sigma_t (x j(x) + x^* j(x^*))
\zeta \rangle \left( f(t-i/4) +f(t+i/4) \right)  \,dt.
\end{eqnarray*}
Thus as a relation between bilinear form on $D(\Delta^{1/4}) \cap
D(\Delta^{-1/4})$, one has
\begin{eqnarray}\label{3.11}
&&\int T(\sigma_t (x j(x) +x^* j(x^*))) f(t)\,dt \\
&& \quad = \int \sigma_t (x j (x) +x^* j(x^*))
(f(t+i/4)+t(t-i/4))\,dt. \nonumber
\end{eqnarray}
Since the linear operators in the above are well-defined bounded
operators by the bound (c) and the fact that $x j(x)+ x^* j(x^* )
\in \cL_{1/4} (\cH)$ and since $D(\Delta^{1/4} ) \cap
D(\Delta^{-1/4})$ is dense in $\cH$, the relation (\ref{3.11})
holds as a relation between bounded operators. It follows from
(\ref{3.9}) and (\ref{3.11}) that
\begin{eqnarray*}
\textmd{II }&=& -\int\langle \xi_+,  \sigma_t (x j (x) +x^*
j(x^*))\xi_-
\rangle  (f(t+i/4)+t(t-i/4))\,dt \\
&\le & 0.
\end{eqnarray*}
Here we have used the property (c) in the Definition \ref{defn2.1}
and the fact that $\sigma_t (Aj(A)) \xi_- = \sigma_t(A)
j(\sigma_t(A)) \xi_- \in \cP$ for any $A \in \cM$. This proved the
property (c).

Clearly $\cE(\xi, \xi) \ge 0$, $ \forall \xi \in \cH$. Thus the
properties (a), (b) and (c) in (\ref{2.1}) hold. Since $\left(
\sigma_{-i/4}(B) -j(\sigma_{-i/4} (B^*)) \right) \xi_0 =0$ for any
$B \in \cM_{1/4}$, $\cE(\xi, \xi_0 ) =0$ for any $\xi \in \cP$.
Thus $\cE$ is a Dirichlet form by Proposition 4.5(b) and
Proposition 4.10 (ii) of \cite{Cip}. $\quad \square$

\section{Proofs of Proposition \ref{*prop2.1} and Theorem
\ref{*thm2.2}}

In this section we prove Proposition \ref{*prop2.1} and Theorem
\ref{*thm2.2}. We first establish  relations equivalent to
(\ref{2.13}) and (\ref{**2.16}) respectively:
\begin{lemma}\label{lemm4.1} (a) For a given $x \in \cM_{1/4}$, the
relation (\ref{2.13}) holds if and only if
$$
\sigma_{i/4}(x)A \sigma_{-i/4}(x^* )=
\sigma_{i/4}(x^*)A\sigma_{-i/4}(x)
$$
for any $A\in \cM$.

(b) For given $\{ x_1, x_2, \cdots, x_n \} \subset \cM_{1/4}$, the
relation (\ref{**2.16}) holds if and only if
$$
\sum_{k=1} ^n \sigma_{i/4} (x_k)A \sigma_{-i/4} (x_k^* )=
\sum_{k=1} ^n \sigma_{i/4} (x_k^*)A \sigma_{-i/4} (x_k )
$$
for any $A \in \cM$.
\end{lemma}
\begin{proof}
(a) By acting $D_{-1/4}$ on the both sides of (\ref{2.13}), it can
be checked that the condition (\ref{2.13}) is equivalent to the
following condition:
$$\sigma_{i/4}(x) j( \sigma_{-i/4}(x)) =
\sigma_{i/4}(x^*) j( \sigma_{-i/4}(x^*)) $$
 Since $j(\sigma_{-i/4}(B)) \xi_0 = \sigma_{-i/4} (B^*) \xi_0$ for any $B\in
\cM_{1/4}$. one has that for any $A \in \cM$
\begin{eqnarray*}
\sigma_{i/4} (x) j(\sigma_{-i/4}(x)) A\xi_0 &= &  \sigma_{i/4}(x) A \sigma_{-i/4} (x^*) \xi_0, \\
 \sigma_{i/4}(x^*) j(\sigma_{-i/4}(x^*)) A\xi_0 &= & \sigma_{i/4}(x^* ) A \sigma_{-i/4} (x) \xi_0.
\end{eqnarray*}
Since $\cM \xi_0 $ is dense,  (\ref{2.13}) holds if and only if
$$
\sigma_{i/4}(x) A\sigma_{-i/4} (x^*)  \xi_0 =\sigma_{i/4}(x^*) A
\sigma_{-i/4} (x) \xi_0.
$$
for any $A \in \cM$. Since $\xi_0$ is a separating vector, we
proved (a).

(b) If one replaces $x$ and $x_k$ and sums over $k$ from 1 to $n$
in the above, the proof of the part (b) follows from that of the
part (a).
\end{proof}

We now turn to the proof of Proposition \ref{*prop2.1}.  Recall
the definitions of the linear maps $D_{1/4}$, $D_{-1/4}, T, S$ and
$I_0$ on $\cL (\cH)$ defined as in (\ref{3.1}) - (\ref{3.4}).

\vspace*{0.3cm} \noindent {\it Proof of Proposition
\ref{*prop2.1}.} Notice that for any $y \in \cM_{1/2}$ and $A\in
\cL_{1/2}(\cH)$ equalities
\begin{eqnarray} \label{4.1}
\Delta^{1/2} A\xi_0 & =& \sigma_{-i/2} (A) \xi_0 , \\
y\xi_0 &=& J \Delta^{1/2} y^* \xi_0 = j(\sigma_{-i/2}(y^*)) \xi_0
 \nonumber
\end{eqnarray}
hold. Let $L$ be given as in (\ref{2.10}). A direct computation
yields
\begin{eqnarray*}
\Delta^{1/4} L(A) \xi_0 &=& \Delta^{1/4} (y^* y - 2y^*
j(\sigma_{-i/2}(y^*)) + j(\sigma_{-i/2} (y^*y))) A\xi_0 \\
&& \,+ \Delta^{1/4} (i Q - i j(\sigma_{-i/2} (Q))) A\xi_0 \\
&=& D_{1/4} (y^* y -2y^* j(\sigma_{-i/2}(y^*))
-j(\sigma_{-i/2}(y^* y))) \sigma_{-i/4} (A) \xi_0 \\
&&\,+i \left( D_{1/4} (Q) -D_{1/4} (j(\sigma_{-i/2}(Q) )) \right)
\sigma_{-i/4} (A) \xi_0
\end{eqnarray*}
Thus it follows from (\ref{2.12}) and the above relation that
\begin{eqnarray*}
H &=& D_{1/4} (y^* y -2y^* j(\sigma_{-i/2}(y^*))
-j(\sigma_{-i/2}(y^* y))) \\
&& \,\,+i  D_{1/4} (Q) -i D_{1/4} (j(\sigma_{-i/2}(Q) )).
\end{eqnarray*}
Since $D_{1/4} (j(\sigma_{-i/2}(B))) = D_{-1/4} (j(B))$  for any
$B\in \cM_{1/2}$, $H$ can be written as
\begin{eqnarray} \label{4.2}
H&=&   D_{1/4} (y^* y)  -2\sigma_{-i/4} (y^*)
j(\sigma_{-i/4}(y^*)) + D_{-1/4} (j(y^* y)) \\
&& \,\,+i  D_{1/4} (Q) -i D_{-1/4} (j(Q) ). \nonumber
\end{eqnarray}
Since $(D_{\pm 1/4} (B))^* = D_{\mp 1/4} (B^*)$ for any $B \in
\cL_{1/2} (\cH)$, one has
\begin{eqnarray}\label{4.3}
H^* &=& D_{-1/4} (y^* y)  -2\sigma_{i/4} (y)
j(\sigma_{i/4}(y)) + D_{1/4} (j(y^* y)) \\
&& \,\,-i  D_{-1/4} (Q) +i D_{1/4} (j(Q) ). \nonumber
\end{eqnarray}
Thus $H =H^*$ if and only if
\begin{eqnarray}\label{4.4}
&& i T(Q) -i T(j(Q)) \\
\nonumber && \,\,\, = - S(y^* y) + 2\sigma_{-i/4}(y^*) j(
\sigma_{-i/4}(y^*)) - 2\sigma_{i/4}(y) j( \sigma_{i/4}(y)) +
S(j(y^* y)).
\end{eqnarray}
If the relation (\ref{2.13}) holds, $H$ is self-adjoint if and
only if
$$ i T(Q) -i T(j(Q)) = -S(y^* y)+ S(j(y^* y)),
$$
and so by Lemma \ref{lemm3.1},
\begin{eqnarray*}
Q-j(Q) &=& i I_0 S(y^* y) -i I_0 S(j(y^* y)) \\
&=& I_0 (i S(y^* y)) - j(I_0 (i S (y^* y))).
\end{eqnarray*}
Here we have used the fact $S(j(y^* y)) = -j (S(y^*y))$. Thus $Q$
can be written as
\begin{equation} \label{4.5}
Q= i I_0 (S(y^* y)) + Q_c
\end{equation}
for some $Q_c \in \mathcal{Z} (\cM)$, where $\mathcal{ Z} (\cM) =
\cM \cap \cM'$. Since $Q_c$ has no contributions in $L$ and $H$,
we take $Q_c =0$. By setting $x:=\sigma_{i/4}(y)$
($y=\sigma_{-i/4}(x)$), the expression of $Q$ in (\ref{2.14})
($Q_c =0$) equals that in (\ref{4.5}).

Next, we substitute (\ref{4.5}) into (\ref{4.2}) to obtain
\begin{eqnarray*}
H&=&  D_{1/4} (y^* y)  -2\sigma_{-i/4} (y^*)
j(\sigma_{-i/4}(y^*)) + D_{-1/4} (j(y^* y)) \\
&& \,\,-  D_{1/4} (I_0 (S(y^*y)) - D_{-1/4} (j(I_0 (S(y^* y)) ).
\end{eqnarray*}
We use Lemma \ref{lemm3.1} and the fact that $I_0 S = SI_0 $ on
$\cL_{1/4} (\cH)$ to conclude that
\begin{eqnarray} \nonumber
H &=& D_{1/4} T(I_0 (y^* y))  -2\sigma_{-i/4} (y^*)
j(\sigma_{-i/4}(y^*)) + D_{-1/4}T(I_0 (j(y^* y)))  \\
\nonumber
&& \,\,-  D_{1/4} S(I_0 (y^*y)) + D_{-1/4} S(I_0(j(y^* y)) )\\
&=& 2I_0 (y^*y) -2 T(I_0 (\sigma_{-i/4}(y^*)
j(\sigma_{-i/4}(y^*))) + 2I_0 (j(y^*y)). \label{4.6}
\end{eqnarray}

Setting $y=\sigma_{-i/4}(x)$ and use (\ref{2.13}). we get
\begin{eqnarray}\nonumber
 &&TI_0 (\sigma_{-i/4}(y^* ) j(\sigma_{-i/4}(y^*)) = I_0 (( D_{1/4} + D_{-1/4}) x^* j(x^*)) \\
\nonumber && \quad\quad = I_0(D_{-1/4} (x^* j(x^*))) + I_0
(D_{1/4}( x
j(x))) \\
\label{4.7} && \quad\quad = I_0 (\sigma_{-i/4} (x)^*
j(\sigma_{-i/4}(x^*))) + I_0 (\sigma_{-i/4}(x)
j((\sigma_{-i/4}(x^*))^*).
\end{eqnarray}
It follows from that Lemma \ref{lemm4.1} (a) that
\begin{equation}\label{4.8}
2y^* y = \sigma_{-i/4}(x)^* \sigma_{-i/4} (x)
+\sigma_{-i/4}(x^*)^* \sigma_{-i/4} (x^*).
\end{equation}
We now substitute the equalities in (\ref{4.7}) and (\ref{4.8})
into  (\ref{4.6}) to conclude that
\begin{eqnarray}\label{4.9}
H&=& I_0 \left( \left[ \sigma_{-i/4}(x)
-j(\sigma_{-i/4}(x^*))\right]^* \left[ \sigma_{-i/4}(x)
-j(\sigma_{-i/4}(x^*))\right] \right) \\
\nonumber && \,\, + I_0 \left( \left[ \sigma_{-i/4}(x^*)
-j(\sigma_{-i/4}(x))\right]^* \left[ \sigma_{-i/4}(x^*)
-j(\sigma_{-i/4}(x))\right] \right).
\end{eqnarray}
By the definition of $I_0$ in (\ref{3.4}), the expression of $H$
in (\ref{4.9}) equals that in (\ref{2.7}) with $f=f_0$. This
proved Proposition \ref{*prop2.1} completely. $\quad \square$

\vspace*{0.3cm} \noindent {\it Proof of Theorem \ref{*thm2.2}.} It
is clear that Theorem \ref{*thm2.2} can be proved by the method
used in the proof of Proposition \ref{*prop2.1}. More precisely,
if one replaces $y$ by $y_k$ and $x$ and $x_k$, respectively, in
every expression containing $y$ and $x$ in the proof of
Proposition \ref{*prop2.1} and sums over $k$ from $1 $ to $n$, and
uses (\ref{**2.16}) and  Lemma \ref{lemm4.1} (b) instead of
(\ref{2.13}) and Lemma \ref{lemm4.1} (a) respectively,  then the
proof of Theorem \ref{*thm2.2} follows from that of Proposition
\ref{*prop2.1}. $\quad \square$.

\section{Discussion}

We have seen that the conditions (\ref{**2.16}) and (\ref{**2.17})
are  sufficient conditions such that the operator $H$ induced by
the Lindblad type map $L$ given as (\ref{**2.15}) is self-adjoint
and the seif-adjoint operator $H$ can be expressed as sum of
Dirichlet operators given by (\ref{2.7}) with $f=f_0$. We would
like to give a brief discussion on the necessary and sufficient
condition for self-adjointness of $H$ and also on the map $L$ on
$\cM$ associated to an Dirichlet operator (\ref{2.7}) for a
general admissible function $f$.

Let $L$ be given as (\ref{2.10}) and let $H$ be the operator on
$\cH$ defined by (\ref{2.12}). The necessary and sufficient
condition for the self-adjointness of $H$ is given by (\ref{4.4})
:
\begin{eqnarray} \label{5.1}
&& i T(Q) - i T( j(Q)) \\
\nonumber && \quad = -S(y^* y) + S(j(y^* y)) + 2 \sigma_{-i/4}(y^*
) j(\sigma_{-i/4}(y^*)) - 2 \sigma_{i/4}(y) j(\sigma_{i/4}(y)).
\end{eqnarray}
From the above relation, one has to express $D_{1/4} (Q)
-D_{-1/4}(j(Q))$ in terms of $y$ and $y^*$, and substitute it into
(\ref{4.2}). In general case, we are not able to estimate $Q$ from
(\ref{5.1}) directly. Thus we have assumed the property
(\ref{2.13}) to estimate $Q$ from (\ref{5.1}).

If the state $\omega$ on $\cM$ defined by $\omega(A) := \langle
\xi_0, A\xi_0 \rangle $ is tracial, i.e., $\omega(AB) =
\omega(BA)$, $\forall A, B \in \cM$, then $\Delta = \1$ and the
relation (\ref{5.1}) becomes
$$
i Q - i j(Q) = y^* j (y^*) - y j(y).
$$
The above relation is equivalent to
\begin{equation} \label{5.2}
i [Q, A] = y^* A y - y A y^*, \quad \forall A \in \cM.
\end{equation}
We substitute (\ref{5.2}) into(\ref{2.10}) and use (\ref{5.2})
again with $A=\1$. Then the map $L$ given in (\ref{2.10}) can be
written as
$$
 L(A) = \frac 12 \{ y^* y A -2 y^* A y + Ay^* y \}
+ \frac 12 \{ yy^* A- 2 y A y^* + Ayy^* \}
$$
If one replace $y$ by $y_k$ in the above argument and sums over
$k$, one can see that the Lindblad type generator (\ref{**2.15})
is symmetric if and only if $L$ can be written as
\begin{eqnarray*}
L(A) &=& \frac 12 \sum_{k=1}^n \{ y_k^* y_k A-2y_k^* A y_k +
Ay_k^* y_k \} \\
&& \,\,\,\, + \frac 12\sum_{k=1}^n \{ y_k y_k^* A-2y_k A y_k^* +
Ay_k y_k^* \}, \quad A \in \cM.
\end{eqnarray*}
Notice that the condition (\ref{**2.16}) and (\ref{**2.17}) in
Theorem \ref{*thm2.2} are satisfied automatically for the map $L$
in the above. Thus if $\xi_0$ defines a tracial state, the
condition (\ref{**2.16}) and the condition (\ref{**2.17}) ($Q=0$)
are also necessary conditions for the map $L$ in (\ref{**2.15}) to
be symmetric, or equivalently the operator $H$ induced by $L$ to
be self-adjoint.

Next, consider a Dirichlet operator (\ref{2.7}) for a given $x \in
\cM_{1/2}$ and an admissible function $f$ in the sense of
Definition \ref{defn2.1}. Let $L : \cM \to \cM$ be the map given
by
\begin{equation} \label{5.3}
L(A) = L^{(1)} (A) + L^{(2)} (A),
\end{equation}
where
\begin{eqnarray}\nonumber
L^{(1)}(A) &=& \frac 12 \int  \Big\{ \sigma_{t+i/4}(x^*)
\sigma_{t-i/4}(x^*)A - 2\sigma_{t+i/4}(x^*)A \sigma_{t-i/4}(x)
\\ \nonumber && \quad \qquad \,\, + A
\sigma_{t+i/4}(x^*)\sigma_{t-i/4}(x) \Big\}
 \left( f(t-i/4) +
f(t+i/4)\right ) \,dt \\ \label{5.4}
  && \quad \,\,\, + \frac i 2 [ Q^{(1)}, A ], \\[0.2cm]
\nonumber  Q^{(1)} (A) &:=& i \int \left \{ \sigma_t (x^* )
\sigma_{t-i/2}(x) -\sigma_{t+i/2}(x^*) \sigma_t (x) \right \} f(t)
\, dt,
\end{eqnarray}
and $L^{(2)} (A)$ is defined as $L^{(1)} (A)$ replacing $x$ by
$x^*$.  Using the fact that for any $ A\in \cL _{1/4} (\cH)$
$$
T \left( \int \sigma_t(A) f(t) \,dt \right) = \int \sigma_t(A)
\left( f(t-i/4) + f(t+i/4)\right) \,dt,
$$
and the method in the proof of Proposition \ref{*prop2.1}, one can
show that $L$ in (\ref{5.3}) and $H$ in (\ref{2.7}) is related by
(\ref{2.12}). We leave the detailed proof to the reader.

\vspace{0.2cm} \noindent {\bf Acknowledgements}:  This work was
supported by Korea Research Foundation (KRF-2003-005-C00010),
Korean Ministry of Education.

\end{document}